\documentclass[aps,prl,twocolumn,showpacs]{revtex4}
\usepackage{amsmath}
\usepackage{amssymb}
\usepackage{graphics}

\newcommand\ket[1]{\left|#1\right>}
\newcommand\bra[1]{\left<#1\right|}

\newcommand\expect[1]{\left<#1\right>}

\newcommand\si{\sigma}
\newcommand\Ev{\mathbb{E}}

\newcommand\dg{^\dagger}

\newcommand\half{\frac{1}{2}}

\begin{document}
\title{Comment on 'Uniqueness of the Equation for Quantum State Vector Collapse'}   
\author{Lajos Di\'osi}
\email{diosi.lajos@wigner.mta.hu}
\homepage{www.rmki.kfki.hu/~diosi}
\affiliation{Wigner Research Center for Physics, 
H-1525 Budapest 114. P.O.Box 49, Hungary}
\date{\today}
\begin{abstract}
A diffusive stochastic Schr\"odinger equation (SSE) is shown for the first time, such that contributes to 
a non-completely positive dynamics. This contradicts to a recent Letter claiming that  SSEs, under most 
general conditions, enforce complete-positivity.  The general form and parametrization of the SSE in the 
Letter is different from an alternative simpler result, the difference is shown to be completely redundant 
because of the gauge-freedom of the state vector's phase.
\end{abstract}
\maketitle
A recent Letter \cite{BasDurHin13} investigated markovian stochastic Schr\"odinger equations (SSEs) under 
the assumption of no-faster-than-light signalling \cite{Gis89}.  I found that Theorem 1, claiming that the evolution 
of the density matrix $\rho$ must be completely-positive (CP), is incorrect. Theorem 2 constructs the most general 
diffusive SSE for the wave function $\psi$,  which looks different from the simpler results in Ref. \cite{WisDio01}.
I prove that the difference is redundant. 

If Theorem 1 were true, no markovian SSE would exist for the non-CP qubit master equation \cite{NCP}:
\begin{equation}\label{drho}
\frac{d\rho}{dt}=\sum_{k=1}^3 c_k \left(\si_k\rho\si_k-\rho\right),
~~~c_1=c_2=-c_3=1.
\end{equation}
I consider the following SSE (cf. \cite{Dio86} for a jump process):
\begin{equation}\label{dpsi}
d\psi = -\half\sum_{k=1}^3 c_k \left(\si_k-n_k\right)^2\psi dt
        +\sqrt{2}n_z \psi_\perp dW
\end{equation}
where $n_k=\bra{\psi}\si_k\ket{\psi}$ and $\psi_\perp$ is orthogonal to $\psi$,
we can express it by $\psi_\perp=(1-n_z^2)^{-1/2}(n_y\si_x-n_x\si_y)\psi$.
The SSE (\ref{dpsi}) yields the master equation (\ref{drho}) for
$\rho=\Ev\ket{\psi}\bra{\psi}$. The proof goes like this. From Eq. (\ref{dpsi}) we get
\begin{equation}\label{pro1}
\frac{d\rho}{dt}=-\half\Ev\sum_{k=1}^3 c_k \left\{\left(\si_k-n_k\right)^2,\ket{\psi}\bra{\psi}\right\}
                 +2\Ev n_z^2\ket{\psi_\perp}\bra{\psi_\perp}.
\end{equation}
One can confirm the identity
\begin{equation}\label{pro2}
2 n_z^2\ket{\psi_\perp}\bra{\psi_\perp}=
\sum_{k=1}^3 c_k \left(\si_k-n_k\right)\ket{\psi}\bra{\psi}\left(\si_k-n_k\right)
\end{equation}
which, when inserted into (\ref{pro1}), leads to the linear master equation (\ref{drho}).  
Hence, Theorem 1 cannot be correct. The proof fails clearly if the number $n$ of  independent 
Lindblad operators $L_k$ is bigger than the dimension $d$ \cite{BasDurHin14}. 
 
For CP master equations, the Letter's Theorem 2  is correct.
The authors mention that Ref. \cite{WisDio01} had answered the same question 
but the Letter does not compare the results. I remedy the omission.
An additional gauge transformation $\psi\rightarrow\exp(-id\chi)\psi$ with phase 
$d\chi=\mathrm{Im}\sum_k\bra{\psi}L^{(\psi)}_k\ket{\psi}(\ell_k^{(\psi)} dt+dW_k)$ 
brings the Letter's SSE (4) to the form 
\begin{align}\label{dpsiWD}
d\psi =\Bigl[&-iHdt+\sum_{k=1}^N\sum_{j=1}^n u_{kj}^{(\psi)}(L_j-\expect{L_j})dW_k\nonumber\\
             &-\half\sum_{k=1}^n(L_k\dg L_k-2\expect{L_k}^\star L_k+\vert\expect{L_k}\vert^2)dt\Bigr]\psi
\end{align}
where $\expect{L_k}=\bra{\psi}L_k\ket{\psi}$. The  matrix $u$ has gone from the drift part! 
The resulting SSE coincides exactly with Eq. (8.1) in Ref. \cite{WisDio01}, implying the 
following relationship between the noises of \cite{WisDio01} and the Letter, respectively:     
\begin{equation}\label{xiW}
d\xi^\ast_j=\sum_{k=1}^N dW_k u_{kj},~~~j=1,2,\dots,n\leq N.
\end{equation} 
In Ref. \cite{WisDio01}, all physically different SSEs are uniquely parametrized by the $n\times n$ 
complex symmetric correlation matrices $s_{jl}=(\mathbb{E}d\xi_jd\xi_l)/dt$ 
(to avoid confusion, here we use $s$ for $u$ of (4.1) in \cite{WisDio01}). 
Now Eq. (\ref{xiW}) establishes the correspondence between the $u$ and $s$:
\begin{equation}\label{suu}
s_{jl}^\ast=\sum_{k=1}^N u_{kj}u_{kl},~~~~j,l=1,2,\dots,n\leq N.
\end{equation}
As I said, the matrix  $s_{jl}$, only constrained by $\Vert s\Vert$, cf. (4.3) in \cite{WisDio01}, 
is in one-to-one correspondence with the  physically different SSEs at a given CP-evolution of $\rho$.
The  matrix $u_{kj}$ is not, its  part  $N\geq j>n$ is redundant. Now (\ref{suu}) shows a further redundancy: 
both $u$ and $Ou$, with any $N\times N$ orthogonal  matrix $O$, yield the same SSE.

Ref.~\cite{WisDio01} derived the SSEs under CP master equation 
from standard quantum monitoring. The SSE (\ref{dpsi}) is the first
diffusive SSE considered ever that underlies non-CP master equation,
its physical relevance, if any,  needs further studies.

This work was supported by the Hungarian Scientific Research Fund (Grant No. 75129)
and the EU COST Action MP1006.

\end{document}